# Phoenix-Chess strategy or revisiting the algorithm for playing in Chess with incomplete information

**Sergey Ershkov***, Affiliation: Plekhanov Russian University of Economics,

Scopus number 60030998, 36 Stremyanny Lane, Moscow, 117997, Russia,

Corresponding author, e-mail: sergej-ershkov@yandex.ru,

**Millana Ershkova**, flat 44, building 24, 60 let VLKSM Street, Zheleznogorsk,

Krasnoyarsk Territory 662978, Russia, e-mail: academic-mail@yandex.ru,

**Abstract**

We present here the new insight or revisiting the algorithm for playing in Chess with incomplete information (which can be recognized by its newly short-name as 'Phoenix-Chess' strategy). The only difference with respect to the classical variant of Chess-game is that each rook after its having been captured by enemy's chess piece in the proccess of gaming is not to be eliminated from the current game, but this rook is assumed being under virtual repairing during next N-steps (the required number of N is discussed in the current research). Then afterwards, such rook will be introduced in game again during *maximal* N-steps if only the chessboard's square (on which it was captured previously) has not been occupied at previous step. In this case, 'Phoenix-Chess' can be classified as game without predictable horizon of planning, so this kind of game should be considered as Chess-like games with incomplete information.

## Introduction.

Having been known as game to which a lot of people have been playing previously during a many thousands of years, Chess is mathematically classified as a two person zero-sum game (this means that there should be two results: one possibility concerns the situation when one person wins, whereas another loses; rather rare case is that of final 'draw' result for the game where both persons have not made a fatal mistake during their game-competition [1]). Nevertheless, we can say from rather theoretical point of view that draws are rare, but they happen frequently in chess real game process. Also from a game-theoretical viewpoint draws may occur more frequently than winning strategies for a side, depending on the domain (tic-tac-toe for example) There are a lot of outstanding fundamental works concerning theoretical findings with respect to Chess in theory of games (for example [2-6], but are not limited to). But most important to note that Chess is also a game of total information: both players are assumed to have complete information about what is happening on chessboard and the location of all the pieces there. In addition, Chess is a finite, but highly combinatorial game (where final result is too hard to be calculated by hand inside the human minds) in which even the top grandmaster can only calculate 20-30 of chess pieces's moves ahead; in a real game, they rely on much intuition rather than search. On average there are 35-40 moves a players can play from both sides of chessboard (either white or black) during a game.

## 1. Phoenix-Chess strategy or revisiting the algorithm for playing in Chess with incomplete information.

Let us present here the algorithm for playing in Chess with incomplete information (which can be recognized by its newly short-name as 'Phoenix-Chess' strategy). The only difference with respect to the classical variant of Chess-game is that each rook after its having been captured by enemy's chess piece in the proccess of game is not to be eliminated from the current game, but this rook is assumed being under virtual repairing during next N-steps (the required number of N is discussed here and below in the current research). Then afterwards, such rook will be introduced in game again at the same place after N-steps if only the chessboard's square (on which it was captured previously) has not been occupied at previous step. If chessboard's square (on which rook was captured previously and where it should appear again) is occupied at the current step, this rook appears on the aforementioned chessboard's square as soon as it will be free.

As mentioned above in the Introduction, on average there are 35-40 moves a players can play from both sides of chessboard. So, from one hand, assumption N = 40 will be senseless for the changing the final result of the game (while locking the rook for 100 moves may still result in it victorious returning to the game in a long play). But from other hand, assumimg N = 3 will be useless from point of view with respect to estimation (by each of players) the possible risk for attacking the rook: indeed, some chess piece could be captured irreversebally during such action (as a result of attacking the rook), but the rook will return to the game again during merely 3 steps later. It seems not reasonable in this case to attack the rook at all (the only case is avoiding the sudden final lose situation; but you may want to capture the rook to continue onwards). Thus, the most realistic value is N = 5 or even N = 10 (we do not consider 'castling move' here and below). Indeed, each chess piece can be associated with variety of possibilities to realize this or that attack to enemy's chess pieces or defence from being under attack by them. Such variety is changing all the time during a game, the possibilities to realize or to convert the successful *chances* into the real success mostly disappear step-by-step during a game (it depends on the experience and game-skills of each of the players). When N is variable, the rook will may reappear at the beginning of the turn of the player that lost it. If we regard N=5 and N=10 as realistic values, meaning that the rook may also appear just before the opponents turn, possibly being directly captured again. There are a lot of variant we can follow in choosing this or that possibilities within aforeformulated rules of chess-like game. Important remarks: re-appearing rook gives you no additional castling rights; explanation (clarification) what happens when the square is occupied: the rook will re-appear when the square is vacated.

The open question is apparently "how many steps should such rook be out of game to not disturb *Nash equlibria* [3-4] with respect to the final result? (if N is variable, but each time should be less than maximal agreed value, say N = 5)" (Fig.1).

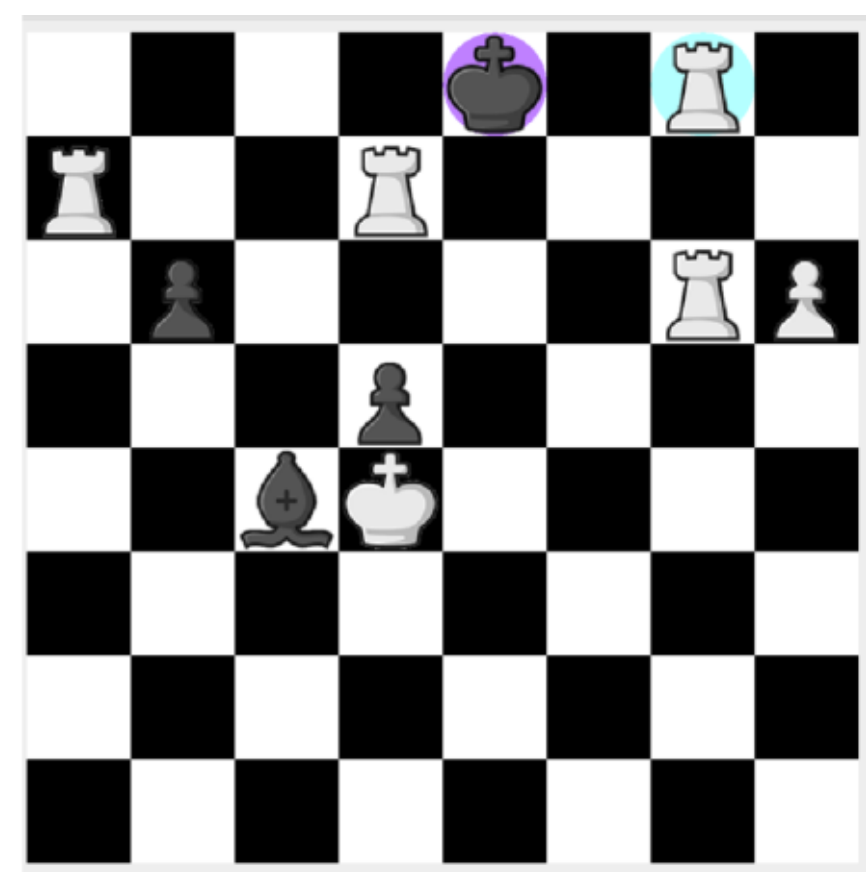

Fig. 1. Schematic imagine for cascade of appearance of 4 rooks on chessboard (step-by-step) after their having been captured on previous steps of game to finalize game by sudden total winning instead of worst final (in classic case of Chess).

Nevertheless, despite of very important theoretical question of existence of the *Nash equlibria* in the aforedescribed 'Phoenix-Chess' strategy (Chess-like games may have no uniform *Nash equilibria* in pure positional strategies [2]), we would suggest *maximal* 5 steps for the above "frozen-stage" for captured rook (as first approximation, i.e. *maximal* N = 5). In this case, 'Phoenix-Chess' can be classified as game without predictable horizon of planning (more than 5 steps) which would be computationally feasible, so this kind of game should be considered as Chess-like games with incomplete information. We will discuss such the realistic evalution with respect to 'Phoenix-Chess' in the next Section.

## 2. Discussion.

Indeed, let us assume for the aforementioned strategy of game with incomplete information that one of players decide (during next 7 steps) to attack the rook of another player. In this case, during first 6 of these 7 steps such rook will be introduced in game again if only the chessboard's square (on which it was captured previously) has not been occupied at previous step.

But we can not exclude the situation that at step N = 6 one of players could decide to sacrifice the useless rook (for some strategical reasons), so the next step of planning for his enemy on chessboard will be senseless.

Thus, the real predictable horizon of planning will be less than 5 steps as we can see from the obvious counter-example as suggested above.

As a side-note remark, it is worthnoting to mention that the Deep Blue won against Garry Kasparov in 1997. Nevertheless, we believe that human thinking process is capable to overcome in a virtual game-battle even the complicated chess software (or other up-to-date sofware or even AI, artificial intelligence) which could do nothing to definitely win following by the suggested 'Phoenix-Chess' algorithm.

Finally, we have developed a simple testing software which has been trained for the aforementioned 'Phoenix-Chess' algorithm with help of AI via updated open-code (this is an in-house program/java-script written). We provide here appropriate examples or results of actual games which stem from the rules suggested in Phoenix-Chess: we schematically imagine them on Figs.2-5 (with approriate comments).

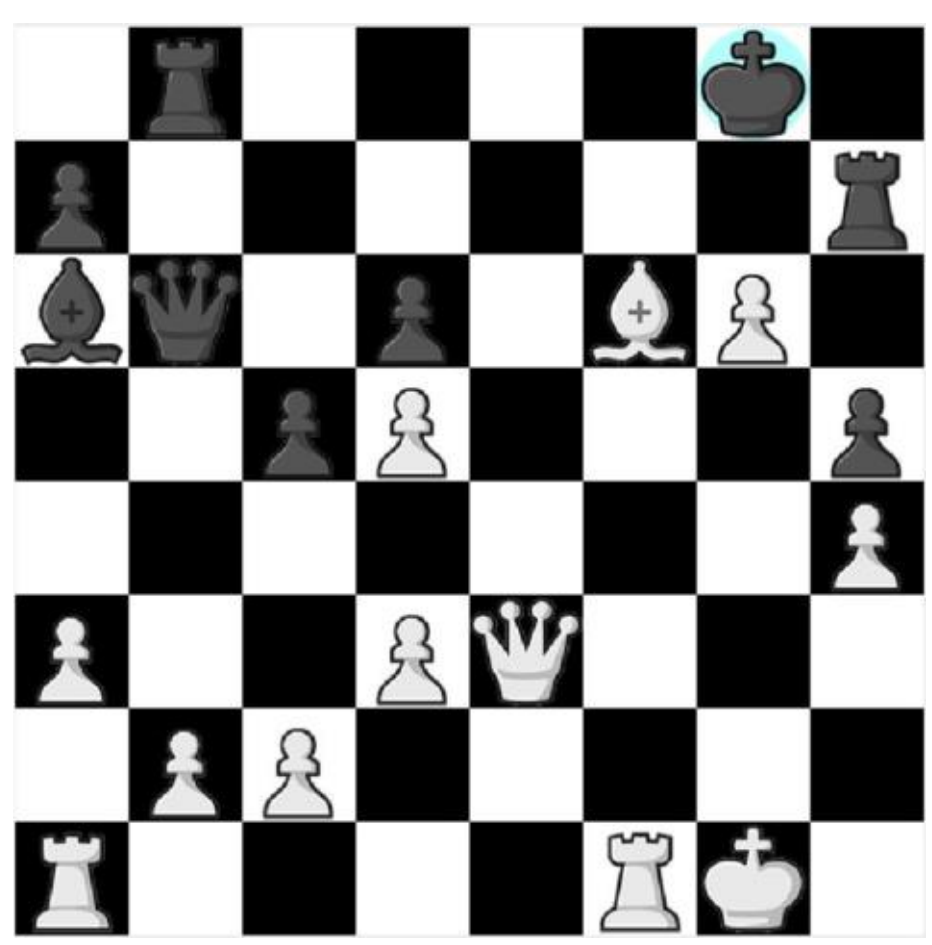

Fig. 2. Initial position. King on G8 has finished the previous step, suggesting pawn on G6 to attack the black rook on H7 with aim of future appearance of such the rook (after capturing) on H7 again, during next *maximal* 3 steps.

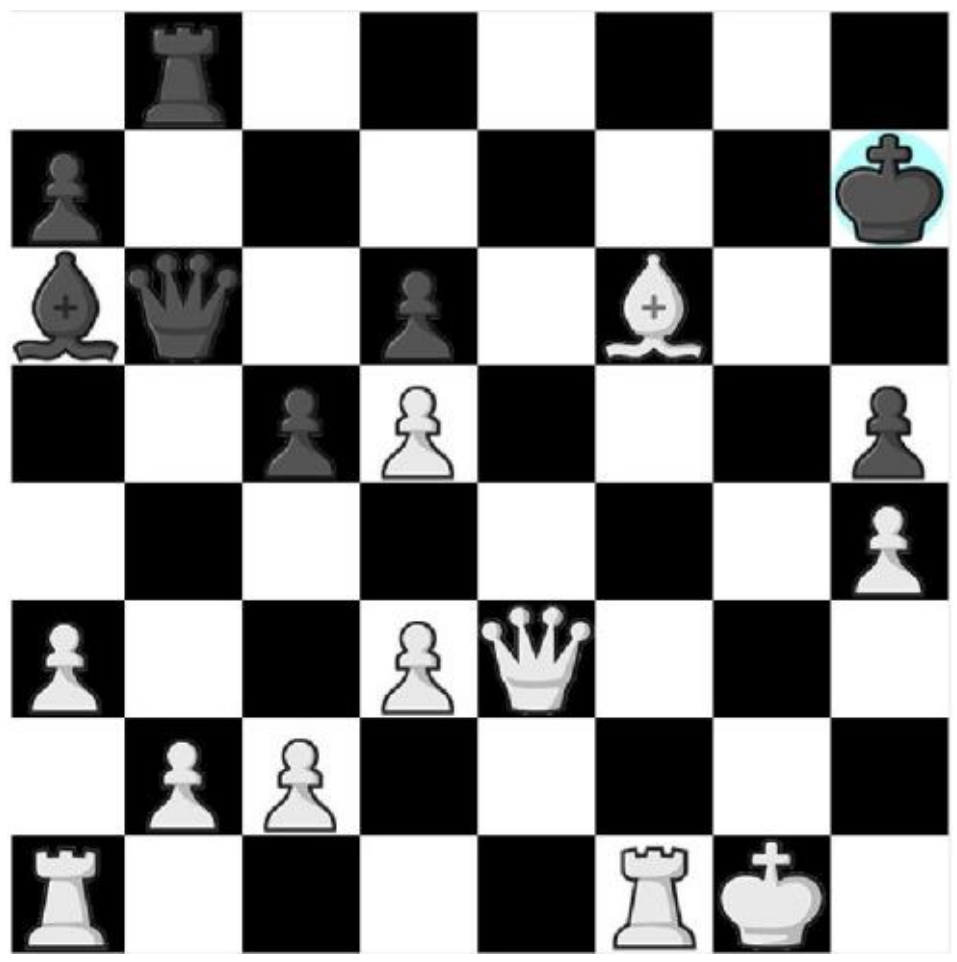

Fig. 3. We present here the next step for the local tactic battle on Fig.2 above.

Pawn from G6 has attacked (and captured) the black rook on H7, then black king has captured this pawn on H7.

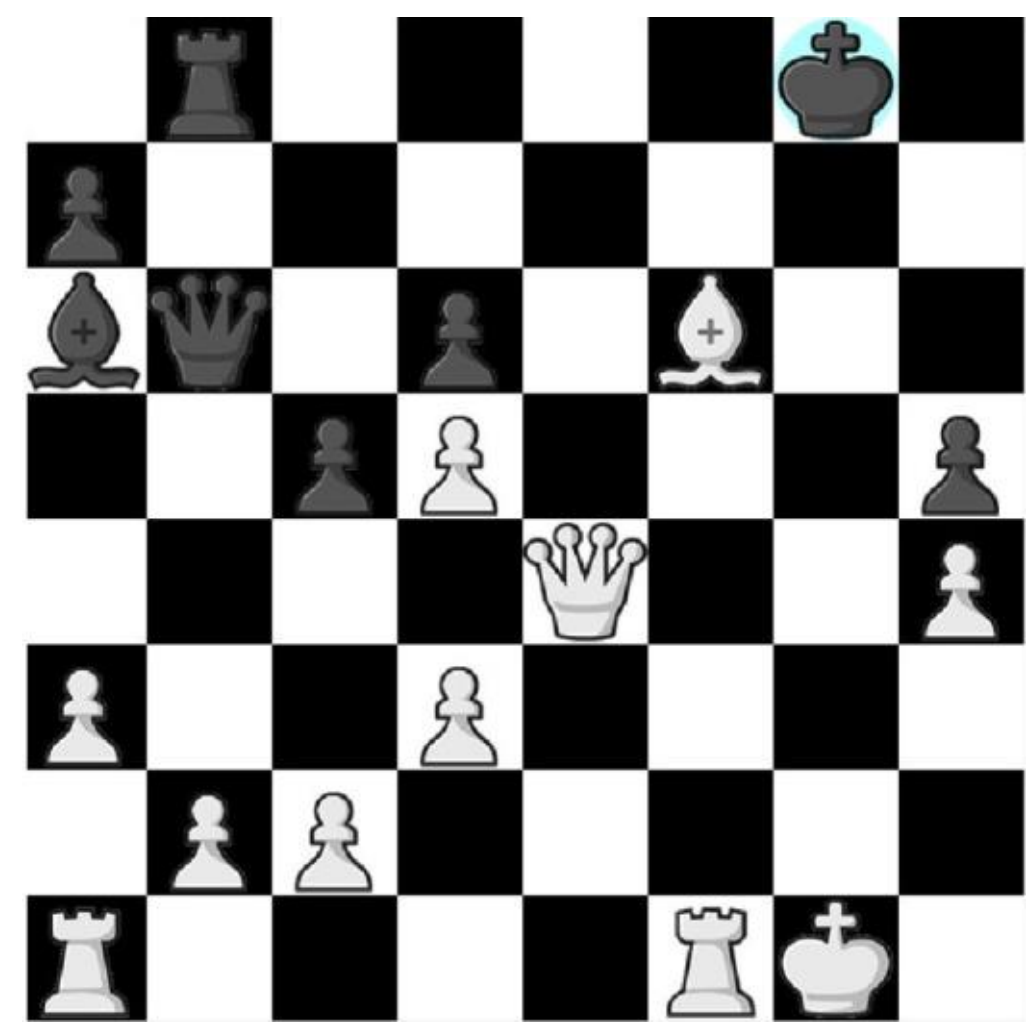

Fig. 4. The next step for the local tactic battle on Fig.3 above. Queen has moved from E3 to E4 with aim to attack the black king, then black king has moved from H7 to G8.

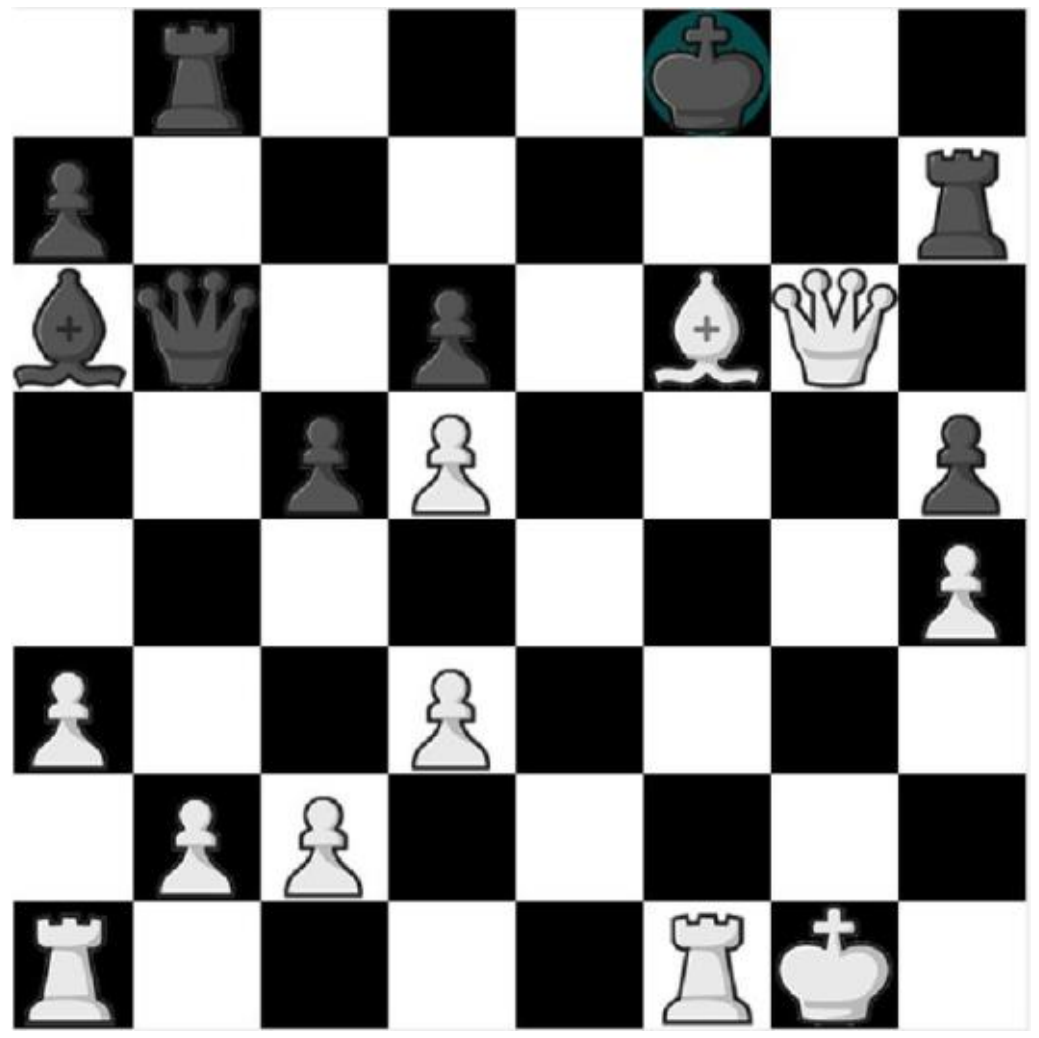

Fig. 5. The next step for the local tactic battle on Fig.4 above. Queen has moved from E4 to G6 with aim to attack the black king, then black king has moved from G8 to F8. But we can see that black rook appears on H7 again.

We should note that we have tested software with *maximal* N =3 for the above "frozen-stage" for rook (as very first approximation). Besides, we preliminary ignore in our soft-ware (just for simplicity) the rule regarding that captured rook will be introduced in game again at the same place after N-steps if only the chessboard's square (on which it was captured previously) has not been occupied at previous step. Nevertheless, we have incoded there the rule that if chessboard's square (on which rook was captured previously and where it should appear again) is occupied at the current step, this rook appears on the aforementioned chessboard's square as soon as it will be free.

## 3. **Conclusion.**

We have presented here the new insight with respect to the algorithm for playing in Chess with incomplete information (which can be recognized by its newly short-name as 'Phoenix-Chess' strategy).

The only difference with respect to the classical variant of Chess-game is that each rook after its having been captured by enemy chess piece in the proccess of game is not to be eliminated from the current game, but this rook is assumed being under *virtual* repairing during next N-steps (the required number of N has been discussed in the current research). Thenafterwards, such rook will be introduced in game again during *maximal* N-steps if only the chessboard's square (on which it was captured previously) has not been occupied at previous step.

The significance of this paper is to prove that aforementioned algorithm for playing Chess-like games of the suggested type does exist, although with certain restrictions as mentioned above. It can be expected that further investigations for the presented 'Phoenix-Chess' strategy can be done.

To make this kind of Chess-like games to be of finite-type, we should suggest rule of "avoiding useless 50 actions or steps in game (from both sides of players)".

The open question should be discussed additionally by professionals who could be interested in introducing a new type of Chess-like game: does the clock of *maximal* N-steps always re-start whenever the square in question is left by the enemy piece, or does the rook never recover when the square is visited again? In case of unlimited recovering of all the rooks during a game, both players theoretically should gain the chance to battle in a combination "king + 4 rooks" against to each other, so those will win who faster finalize game by total winning (see Fig.1).

Most importantly, we should note that the suggested game does have perfect but incomplete information - both players can observe and count the moves and know when and if the rook can be re-introduced during *maximal* N-steps. Despite the fact that known strategies of playing the chess could not be applied in this game to predict the making decisions by players because they simply may use (or not) the strategy of occupying the square where rook was captured before its future recovering during next *maximal* N-steps (with aim to prevent recovering of such the rook).

This is not apparent from the board position alone and requires observing the history of play, but the same applies to decide whether castling is allowed. An obvious example of Chess-like game with imperfect information is a game where player's don't know the state of the game, as in Kriegsspiel.

Even agreeing with obvious self-criticism as discussed above regarding that the suggested game does have perfect but incomplete information, we should note that a simple updating to the rules may make such problem to be obviously solved: namely, we could state that the captured rook should be return back to the game not strictly *after* N-steps, but *during* N-steps (i.e., it may be e.g. (N-2)-step or (N-1) step). In this case, there are no chances to trace all the variety of possible decisions planning by each player how they will move their pieces on chess-desk. One of popular possible strategy could be e.g. the defending of the square where enemy's rook was captured (to prevent its future recovering during *maximal* N-steps). It is worth to note that even in games that actually have incomplete information you can plan ahead (by using chance nodes in a tree search, for example).

The last but not least, the references [2] and [6] refer to games on graphs, where the term "positional" refers to the graph structure and "Chess-like" is a term in the aforementioned papers that does not have much to do with Chess. Theese researches are the examples of profound works where *Nash equilibria* has theoretically been discussed. See also the profound review where authors analyzes how to design engaging and balanced sets of game rules. They concluded that modern chess has evolved over centuries, but without a similar recourse to history, the consequences of rule changes to game dynamics are difficult to predict. Moreover, they introduced their intellectual software product 'AlphaZero' to creatively explore and design new chess variants because there is growing interest in chess variants like Fischer Random Chess, because of classical chess's voluminous opening theory, the high percentage of draws in professional play, and the non-negligible number of games that end while both players are still in their home preparation. Authors compared in [7] nine other variants that involve atomic changes to the rules of chess. The changes allow for novel strategic and tactical patterns to emerge, while keeping the games close to the original. Their findings demonstrate the rich possibilities that lie beyond the rules of modern chess.

Finalizing our short note on topic of introducing the updating with respect to the classical rules of playing chess which should breathe a new life and make a proccess of this old-known game very interesting and unpredictable, we should especially note that 'Phoenix-Chess' can be classified as game without predictable horizon of planning, so this kind of game should be considered as Chess-like games with incomplete information.

## Declarations

### Ethical Approval

Not applicable.

### Competing interests

The Authors declare that there is no conflict of interest or competing interest.

## Authors' contributions

In this research, Dr. Sergey Ershkov is responsible for the results of the article, the formulating of algorithm, simple logical manipulations, conclusions, the representation of a general ansatz and presentation of graphical current or final results of the games.

Millana Ershkova is responsible for inventing and suggesting a general idea for this research, as well as having contributed by playing in a lot of practical games on software which has been trained for the aforementioned ‘Phoenix-Chess’ algorithm with help of AI via updated open-code, also she helped in preparing Figs. 1-5 for the current research.

## Funding

This research received no specific grant from any funding agency in the public, commercial, or not-for-profit sectors.

## Availability of data and materials

The data for this paper are available by contacting the corresponding author.

## Figures

Figs. 1-5 extracted from an in-house program/java-script written with help of AI via updated open-code.